\documentclass[preprint,aps,prd,superscriptaddress]{revtex4}
\usepackage{graphicx,amsfonts,amsmath}
\usepackage{dcolumn}
\usepackage{bm}

\newcommand{\be}{\begin{equation}}
\newcommand{\ee}{\end{equation}}
\newcommand{\ba}{\begin{eqnarray}}
\newcommand{\ea}{\end{eqnarray}}

\newcommand{\Slash}[1]{{#1}\!\!/}
\newcommand{\SLASH}[1]{{#1}\!\!\!/}
\newcommand{\RM}[1]{\mathrm{#1}}

\begin{document}

\title{Lorentz violation and the proper-time method}

\author{T. Mariz}
\affiliation{Instituto de F\'\i sica, Universidade de S\~ao Paulo\\
Caixa Postal 66318, 05315-970, S\~ao Paulo, SP, Brazil}
\email{tmariz,jroberto,liner,ajsilva@fma.if.usp.br}

\author{J. R. Nascimento} 
\affiliation{Instituto de F\'\i sica, Universidade de S\~ao Paulo\\
Caixa Postal 66318, 05315-970, S\~ao Paulo, SP, Brazil}
\affiliation{Departamento de F\'{\i}sica, Universidade Federal da Para\'{\i}ba\\
 Caixa Postal 5008, 58051-970, Jo\~ao Pessoa, Para\'{\i}ba, Brazil}
\email{jroberto, petrov@fisica.ufpb.br}

\author{A. Yu. Petrov}
\affiliation{Departamento de F\'{\i}sica, Universidade Federal da Para\'{\i}ba\\
 Caixa Postal 5008, 58051-970, Jo\~ao Pessoa, Para\'{\i}ba, Brazil}

\author{L. Y. Santos}
\affiliation{Instituto de F\'\i sica, Universidade de S\~ao Paulo\\
Caixa Postal 66318, 05315-970, S\~ao Paulo, SP, Brazil}

\author{A. J. da Silva}
\affiliation{Instituto de F\'\i sica, Universidade de S\~ao Paulo\\
Caixa Postal 66318, 05315-970, S\~ao Paulo, SP, Brazil}

\begin{abstract}
In this paper, we apply the proper-time method to generate the Lorentz-violating Chern-Simons terms in the four-dimensional Yang-Mills and non-linearized gravity theories. It is shown that the coefficient of the induced Chern-Simons term is finite but regularization dependent.
\end{abstract}

\maketitle
\section{Introduction}

The possibility of Lorentz symmetry breaking was firstly suggested in \cite{KosSam,CarFieJac,ColKos} and has motivations from the GZK effect \cite{GZK}, quantum gravity problems \cite{Ame}, and the concept of double special relativity \cite{Mag}. One of the implications of the Lorentz symmetry breaking is the possibility of arising of new classes of couplings in the Lagrangian which involve constant vectors or tensors, similarly to the Seiberg-Witten map \cite{SeiWit} representation of the noncommutative field theories in which the Lorentz-symmetry breaking can be naturally treated as an implication of the space-time noncommutativity \cite{CarKos}. Following the common methodology, the Lorentz-breaking terms in different field theories arise as radiative corrections generated from the coupling of the dynamical fields to spinor fields, which are also coupled with constant vectors or tensors. The most important results achieved in such direction are the generation of the four-dimensional Lorentz-breaking Chern-Simons term in the electrodynamics, both at zero \cite{1,2,3,4,5,6,7,8,9,10,11,12,13,14,15,16} and at a finite temperature \cite{Nas,Cer,Ebe,Mar,Nas2}, in the Yang-Mills theory \cite{Gom}, and the generation of the linearized gravitational Chern-Simons term \cite{Mar2}. 

The characteristic feature of the Lorentz-breaking theories is the ambiguity of quantum corrections, which in the vector field case is related to the presence of the ABJ anomaly \cite{Jac} and has been intensively discussed in the papers \cite{Gom,Jac,Per}. Thus, the natural question is whether the analogous ambiguity takes place in the case of the gravity theory. In particular, the problem is whether this ambiguity survives within the proper-time method \cite{Sch}, which seems to be the most adequate for calculations in the non-linearized gravity (see \cite{Oji} for the calculations of the three-dimensional gravitational Chern-Simons term) since it is known to preserve the gauge invariance and obtain explicit solutions for the equations of motion. At the same time, it is known (see \cite{10,11,12,14}) that even within the framework of the proper-time approach the ambiguity of results is observed in the case of the electrodynamics. So, it is interesting to compare this situation with the gravity case. 

In this paper we apply the proper-time method to find the four-dimensional Lorentz-breaking Chern-Simons terms in the Yang-Mills theory \cite{ColMcD} and the non-linearized gravity \cite{JacPi}. We note that in the case of gravity it is natural to expect that the possible ambiguities are related to the gravitational triangle anomaly \cite{AlvWit} which could be treated as the natural gravitational analogs of the well-known ABJ triangle anomaly. 

The structure of the paper looks as follows. In the next section we calculate the Chern-Simons term in the four-dimensional Yang-Mills theory via the proper-time method. The section III is devoted to the calculation of the Chern-Simons term in the non-linearized gravity within the framework of the same method and the discussion of the possible ambiguities. In the Summary, a review of the results obtained is given. 

\section{Induced Non-Abelian Chern-Simons term}

The starting point of our study is the action of the spinors coupled to the Yang-Mills field with the inclusion of a Lorentz-breaking term proportional to the constant vector $b^\mu$, given by
\be
S=\int d^4x\,\bar{\psi}\left(i\SLASH{\partial}-m-\Slash{b}\gamma_5-g\gamma^\mu A_\mu^aT^a\right)\psi,
\ee
where $T^a$ are the generators of some Lie group algebra satisfying the relations $\RM{tr}(T^aT^b)=\delta^{ab}$ and $[T^a,T^b]=if^{abc}T^c$, and $A_\mu=A_\mu^aT^a$ is the Lie-algebra valued Yang-Mills vector field. The one-loop effective action of the $A_\mu$, obtained via integration over the fermions in this action, can be expressed in terms of the functional trace as \cite{Gom}
\be\label{Seff}
S_\RM{eff}=-i\RM{Tr}\ln(i\SLASH{\partial}-m-\Slash{b}\gamma_5-g\SLASH{A}),
\ee
where $\RM{Tr}$ means trace over Dirac matrices, over the group indices, as well as trace over the integration in
momentum or coordinate spaces. In the sequel we shall use the notation $\SLASH{A}=\gamma^\mu A_\mu^aT^a$. Now, to apply the proper-time method, analogous to the one used to obtain the consistent anomalies \cite{AlvGin}, to calculate this trace we add to this effective action a constant 
\be\label{cte}
C=-i\RM{Tr}\ln(i\SLASH{\partial}+m+\Slash{b}\gamma_5)
\ee 
(for more details see \cite{Oji}), so that after some manipulations, we get
\be\label{Seffp}
S'_\RM{eff}=-i\RM{Tr}\ln\left[-\Box-ig\SLASH{A}\SLASH{\partial}-mg\SLASH{A}-m^2-(g\SLASH{A}+2m)\Slash{b}\gamma_5+2i(b\cdot\partial)\gamma_5-b^2\right].
\ee
Our aim consists in the calculation of $S'_\RM{eff}$ up to first order in the Lorentz-breaking vector $b^\mu$. Using the relation
\be
\ln(A+B)=\ln A+A^{-1}B+\cdots,
\ee
we can write down the first-order term in $b^\mu$ as
\be\label{gb0}
S_\RM{b}=-i\RM{Tr}\,(\Box+ig\SLASH{A}\SLASH{\partial}+mg\SLASH{A}+m^2)^{-1}\left[(g\SLASH{A}+2m)\Slash{b}\gamma_5-2i(b\cdot\partial)\gamma_5
\right].
\ee
We introduce the proper-time representation 
\be
(\Box+ig\SLASH{A}\SLASH{\partial}+mg\SLASH{A}+m^2)^{-1}=\int_0^{\infty}ds e^{-s\left(\Box+ig\SLASH{A}\SLASH{\partial}+mg\SLASH{A}+m^2\right)},
\ee
and rewrite the expression (\ref{gb0}) as
\be\label{gb}
S_\RM{b}=-i\RM{Tr}\,\int_0^{\infty}ds\,e^{-sm^2} e^{-s\left(\Box+ig\SLASH{A}\SLASH{\partial}+mg\SLASH{A}\right)}\left[(g\SLASH{A}+2m)\Slash{b}\gamma_5-2i(b\cdot\partial)\gamma_5
\right].
\ee
Since we are interested in getting the Chern-Simons term, which have at most one derivative in the $A_\mu$ field, we use the Campbell-Hausdorff-Baker formula to get
\be
e^{-s\left(\Box+ig\SLASH{A}\SLASH{\partial}+mg\SLASH{A}\right)}\simeq e^{-s\Box}e^{-s(ig\SLASH{A}\SLASH{\partial}+mg\SLASH{A})}e^{-\frac{s^2}2[\Box,ig\SLASH{A}\SLASH{\partial}+mg\SLASH{A}]},
\ee
where $[\Box,ig\SLASH{A}\SLASH{\partial}+mg\SLASH{A}] \simeq 2ig(\partial_\alpha\SLASH{A})\SLASH{\partial}\partial^\alpha + 2mg (\partial_\alpha\SLASH{A})\partial^\alpha$, up to irrelevant terms with the second derivative in the $A_\mu$. The result is
\ba
S_\RM{b}&=&-i\RM{Tr}\,\int_0^{\infty}ds\,e^{-sm^2} e^{-s\left(ig\SLASH{A}\SLASH{\partial}+mg\SLASH{A}\right)}e^{-s^2(ig(\partial_\alpha\SLASH{A})\SLASH{\partial} + mg (\partial_\alpha\SLASH{A}))\partial^\alpha}\nonumber\\
&&\times\left[(g\SLASH{A}+2m)\Slash{b}\gamma_5-2i(b\cdot\partial)\gamma_5 \right] e^{-s\Box},
\ea
where we have used the cyclic property of the trace. By expanding the exponentials in this expression up to the third order in $\SLASH{A}$ and up to the first order in $\partial_\alpha\SLASH{A}$, we get
\ba\label{Sb}
S_\RM{b}&=&-i{\rm Tr}\int_0^{\infty}ds\,e^{-sm^2}\left[1-s(ig\SLASH{A}\SLASH{\partial}+mg\SLASH{A})+\frac{s^2}{2}(ig\SLASH{A}\SLASH{\partial}+mg\SLASH{A})^2-\frac{s^3}{6}(ig\SLASH{A}\SLASH{\partial}+mg\SLASH{A})^3\right] \nonumber\\
&&\times \left[1- s^2 ig(\partial_\alpha\SLASH{A})\SLASH{\partial}\partial^\alpha - s^2 mg (\partial_\alpha\SLASH{A})\partial^\alpha\right] \left[(g\SLASH{A}+2m)\Slash{b}\gamma_5-2i(b\cdot\partial)\gamma_5 \right]
e^{-s\Box} + \cdots,
\ea
where here the derivatives act on every functions to its right. By dimensional reasons, only the mass-independent terms can produce UV divergences. Thus, the divergent contribution to the Chern-Simons action proportional to $b_\mu A_\nu\partial_\lambda A_\rho$, after we use the definition of the trace 
\be
\RM{Tr}\,\hat{\cal O} = \RM{tr}_\RM{D}\RM{tr}\int d^4x\langle x|\hat{\cal O}|x'\rangle\big|_{x'=x}=\RM{tr}_\RM{D}\RM{tr}\int d^4x\,{\cal O}\delta(x-x')\big|_{x'=x},
\ee
yields
\ba\label{S2div0}
S^{(2)}_\RM{div} &=& g^2\,\RM{tr}_\RM{D}\RM{tr}\int d^4x\int_0^{\infty}ds\,e^{-sm^2}\left[ -s^2(\partial_\alpha\SLASH{A})\SLASH{\partial}\partial^\alpha\SLASH{A}\Slash{b}\gamma_5+2s^3\SLASH{A}\SLASH{\partial}(\partial_\alpha\SLASH{A})\SLASH{\partial}\partial^\alpha b\cdot\partial\gamma_5\right.\nonumber\\\
&&\left.-s\SLASH{A}(\SLASH{\partial}\SLASH{A})\Slash{b}\gamma_5+s^2\SLASH{A}(\SLASH{\partial}\SLASH{A})\SLASH{\partial}b\cdot\partial\gamma_5\right]e^{-s\Box}\delta(x-x')\big|_{x'=x},
\ea
where $\RM{tr}_\RM{D}$ means trace over Dirac matrices, $\RM{tr}$ means trace over group indices, and, except for the derivatives inside the parenthesis, all the derivatives are applied in the delta function. By taking the trace over Dirac matrices and using the Fourier representation of the delta function, we obtain
\ba\label{S2div}
S^{(2)}_\RM{div} &=& -4ig^2\,\RM{tr}\int d^4x\int_0^{\infty}ds\,e^{-sm^2} b_\mu A_\nu\partial_\lambda A_\rho \\
&&\times\int\frac{d^4k}{(2\pi)^4}e^{sk^2}\left(s^2\epsilon^{\alpha\nu\lambda\rho}k_\alpha k^\mu + s^2\epsilon^{\mu\nu\alpha\rho}k_\alpha k^\lambda + s\epsilon^{\mu\nu\lambda\rho}\right). \nonumber	
\ea
Proceeding in a similar way, we can show that the divergent part proportional to $b_\mu A_\nu A_\lambda A_\rho$ is equal to
\be\label{S3div0}
S^{(3)}_\RM{div}= ig^3\,\RM{tr}_\RM{D}\RM{tr} \int d^4x\!\int_0^{\infty}\!ds\,e^{-sm^2}\!\left[\frac{s^2}{2}\SLASH{A}\SLASH{\partial}\SLASH{A}\SLASH{\partial}\SLASH{A}\Slash{b}\gamma_5-\frac{s^3}{3}\SLASH{A}\SLASH{\partial}\SLASH{A}\SLASH{\partial}\SLASH{A}\SLASH{\partial}(b\cdot\partial)\gamma_5\right]e^{-s\Box}\delta(x-x')\big|_{x'=x},
\ee
so that we have
\ba\label{S3div}
S^{(3)}_\RM{div} &=& 4g^3\,\RM{tr} \int d^4x\int_0^{\infty}ds\,e^{-sm^2} b_\mu A_\nu A_\lambda A_\rho \\ &&\times\int\frac{d^4k}{(2\pi)^4}e^{sk^2}\left(\frac{s^2}2\epsilon^{\mu\nu\lambda\rho}k^2-s^2\epsilon^{\mu\nu\alpha\rho}k_\alpha k^\lambda+\frac{s^3}3\epsilon^{\alpha\nu\lambda\rho}k_\alpha k^\mu k^2 \right). \nonumber 
\ea 
By substituting the following integrals
\ba
\label{ints}
\int\frac{d^4k}{(2\pi)^4}e^{sk^2}&=&\frac{i}{16\pi^2s^2},\nonumber\\
\int\frac{d^4k}{(2\pi)^4}e^{sk^2}k_\mu k_\nu &=&\frac{-i}{32\pi^2s^3}g_{\mu\nu},\nonumber\\
\int\frac{d^4k}{(2\pi)^4}e^{sk^2}k_\mu k_\nu k_\lambda k_\rho &=& \frac{i}{64\pi^2s^4}(g_{\mu\nu}g_{\lambda\rho}+g_{\mu\lambda}g_{\nu\rho}+g_{\mu\rho}g_{\nu\lambda})
\ea
in (\ref{S2div}) and (\ref{S3div}), we can see that the several monomials cancel each other, resulting in $S^{(2)}_\RM{div} = S^{(3)}_\RM{div} = 0$. 

It remains to study the finite part of (\ref{Sb}) that contributes to the Chern-Simons action. The $b_\mu A_\nu\partial_\lambda A_\rho$ finite terms of (\ref{Sb}), after disregarding the zero-trace terms or terms vanishing by symmetric integration, yields
\ba\label{S2fin0}
S^{(2)}_\RM{fin} &=& g^2\,\RM{tr}_\RM{D}\RM{tr} \int d^4x\int_0^{\infty}ds\,e^{-sm^2}\left[ s^2m^2\SLASH{A}(\SLASH{\partial}\SLASH{A})\Slash{b}\gamma_5+2m^2s^3\SLASH{A}\SLASH{\partial}(\partial_\alpha\SLASH{A})\partial^\alpha\Slash{b}\gamma_5\right. \nonumber\\
&&\left.+2m^2s^3\SLASH{A}(\partial_\alpha\SLASH{A})\partial^\alpha\SLASH{\partial}\Slash{b}\gamma_5\right]e^{-s\Box}\delta(x-x')\big|_{x'=x},
\ea
which after the calculation of the trace, gives
\be
S^{(2)}_\RM{fin} = -4ig^2\,\RM{tr} \int d^4x\int_0^{\infty}ds\,e^{-sm^2} b_\mu A_\nu\partial_\lambda A_\rho \int\frac{d^4k}{(2\pi)^4}e^{sk^2}s^2m^2\epsilon^{\mu\nu\lambda\rho}. 
\ee
Now, by using the first expression in (\ref{ints}) and the integration over the parameter $s$
\be\label{int}
\int_0^{\infty}ds\,e^{-sm^2}s^{z-1} = \frac{\Gamma(z)}{m^{2z}},
\ee
we obtain 
\be\label{S2fin}
S^{(2)}_\RM{fin} = -\frac{g^2}{4\pi^2}\RM{tr}\int d^4x	\epsilon^{\mu\nu\lambda\rho}b_\mu A_\nu\partial_\lambda A_\rho.
\ee
Finally, the relevant $b_\mu A_\nu A_\lambda A_\rho$ finite terms of (\ref{Sb}) are given by
\ba\label{S3fin0}
S^{(3)}_\RM{fin} &=& -ig^3 \RM{tr}_\RM{D}\RM{tr}\! \int d^4x\! \int_0^{\infty}ds\,e^{-sm^2}\left[\frac{s^2}{2}m^2\SLASH{A}\SLASH{A}\SLASH{A}\Slash{b}\gamma^5\!+\!\frac{s^3}{3}m^2\left(\SLASH{A}\SLASH{\partial}\SLASH{A}\SLASH{\partial}\SLASH{A}\!+\!\SLASH{A}\SLASH{\partial}\SLASH{A}\SLASH{A}\SLASH{\partial}\!+\!\SLASH{A}\SLASH{A}\SLASH{\partial}\SLASH{A}\SLASH{\partial}\right)\Slash{b}\gamma_5\right.\nonumber\\
&&\left.-\frac{s^3}{3}m^2\left(\SLASH{A}\SLASH{\partial}\SLASH{A}\SLASH{A}+\SLASH{A}\SLASH{A}\SLASH{\partial}\SLASH{A}+\SLASH{A}\SLASH{A}\SLASH{A}\SLASH{\partial}\right)b\cdot\partial\gamma_5-\frac{s^3}{3}m^4\SLASH{A}\SLASH{A}\SLASH{A}\Slash{b}\gamma^5\right]e^{-s\Box}\delta(x-x')\big|_{x'=x}
\ea
or, taking into account the trace of Dirac matrices,
\ba\label{S3fin0a}
S^{(3)}_\RM{fin} &=& -4g^3\,\RM{tr} \int d^4x\int_0^{\infty}ds\,e^{-sm^2} b_\mu A_\nu A_\lambda A_\rho \\ &&\times\int\frac{d^4k}{(2\pi)^4}e^{sk^2}\left(\frac{s^2}2m^2\epsilon^{\mu\nu\lambda\rho}+\frac{s^3}3m^2\epsilon^{\mu\nu\lambda\rho}k^2+\frac{s^3}3m^2\epsilon^{\alpha\nu\lambda\rho}k_\alpha k^\mu-\frac{s^3}3m^4\epsilon^{\mu\nu\lambda\rho}\right). \nonumber
\ea
Thus, by integrating over the momenta $k$ and the parameter $s$, we obtain
\be\label{S3fin}
S^{(3)}_\RM{fin} = \frac{ig^3}{6\pi^2}\RM{tr}\int d^4x \epsilon^{\mu\nu\lambda\rho} b_\mu A_\nu A_\lambda A_\rho.
\ee
Therefore, combining both contributions, Eq. (\ref{S2fin}) and Eq. (\ref{S3fin}), we find the result  
\be
S_\RM{CS} = -\frac{g^2}{4\pi^2}\RM{tr} \int d^4x \epsilon^{\mu\nu\lambda\rho}b_\mu\left(A_\nu\partial_\lambda A_\rho - \frac{2ig}{3}A_\nu A_\lambda A_\rho\right),
\ee
which is exactly the non-abelian Chern-Simons term \cite{ColMcD}. Its coefficient coincides with one of the values gotten in \cite{Gom} where it was calculated in two different regularization schemes.

\section{Induced Gravitational Chern-Simons term}

The action in which we are now interested is the one of the spinors coupled to the gravity with the inclusion of a Lorentz-breaking term proportional to the constant vector $b^\mu$ \cite{Mar2},
\be\label{S1}
S = \int d^4x\,e\,{e^\mu}_a\,\bar\psi\left(iD_\mu\gamma^a-b_\mu \gamma^a\gamma_5\right)\psi,
\ee
where ${e^\mu}_a$ is the tetrad (vierbein), and $e\equiv\det {e^\mu}_a$. The covariant derivative is given by
\be\label{D}
D_\mu\psi = \partial_\mu\psi - \frac{i}{4} \omega_{\mu bc}\sigma^{bc}\psi,
\ee
where ${w_\mu}^{bc}$ is the spin connection and $\sigma^{bc} = \frac i2[\gamma^b, \gamma^c]$. Using these expressions and adding a massive term we can rewrite the Eq. (\ref{S1}) as follows
\be
S = \int d^4x\,e\,\bar\psi\left(i\SLASH{\partial} - m - \Slash{b}\gamma_5 + \SLASH{\omega}\right)\psi,
\ee
where $\gamma^\mu = {e^\mu}_a\gamma^a$ and $\omega_\mu=\frac{1}{4}\omega_{\mu bc}\sigma^{bc}$. The corresponding one-loop effective action of the $\omega_{\mu bc}$ can be expressed as
\ba
S_\RM{eff}=-i\RM{Tr}\ln\left(i\SLASH{\partial} - m - \Slash{b}\gamma_5 + \SLASH{\omega}\right).
\ea
Observe that this expression is similar to Eq. (\ref{Seff}), when we change $\SLASH{\omega} \to -g\SLASH{A}$. So, the effective action becomes
\be
S'_\RM{eff}=-i\RM{Tr}\ln\left[-\Box+i\SLASH{\omega}\SLASH{\partial}+m\SLASH{\omega}-m^2+(\SLASH{\omega}-2m)\Slash{b}\gamma_5+2i(b\cdot\partial)\gamma_5-b^2\right].
\ee
As this equation is also similar to expression (\ref{Seffp}), the divergent and finite contributions are similar to those obtained in the non-abelian case, expressions (\ref{S2div}), (\ref{S3div}), (\ref{S2fin}), and (\ref{S3fin}), respectively. The only difference is in the trace over Dirac matrices due to the presence of the $\sigma^{bc}$ matrices. Nevertheless, as these modifications do not affect the tensorial structure the divergent terms in this case also vanish. From now on, we shall only concentrate in the finite terms from which the gravitational Chern-Simons term must appear.  

Thus, the finite contribution proportional to $b_\mu\partial_\nu\omega_{\lambda ab}\,{\omega_\rho}^{ba}$ is given by
\ba\label{SG2fin}
S^{(2)}_\RM{fin} &=& \RM{tr}_\RM{D} \int d^4x\int_0^{\infty}ds\,e^{-sm^2}\left[ s^2m^2\SLASH{\omega}(\SLASH{\partial}\SLASH{\omega})\Slash{b}\gamma_5+2m^2s^3\SLASH{\omega}\SLASH{\partial}(\partial_\alpha\SLASH{\omega})\partial^\alpha\Slash{b}\gamma_5\right. \nonumber\\
&&\left.+2m^2s^3\SLASH{\omega}(\partial_\alpha\SLASH{\omega})\partial^\alpha\SLASH{\partial}\Slash{b}\gamma_5\right]e^{-s\Box}\delta(x-x')\big|_{x'=x}.
\ea
Here, we have another modification due to the introduction of the geodesic bi-scalar $\sigma(x,x')$ in the delta function as follows \cite{Fuj,FujOji}
\be
\delta(x-x') = \int\frac{d^4k}{(2\pi)^4} e^{ik_\alpha D^\alpha\sigma(x,x')}.
\ee
But as in the limit $x'\to x$ we have
\be
D^\alpha D^\beta\sigma(x,x')\big|_{x'=x} = g^{\alpha\beta},
\ee
it is sufficient to complete the covariant derivatives in (\ref{SG2fin}) by using the expression (\ref{D}), through the substitution $\partial_\alpha = D_\alpha + i\omega_\alpha$. We get
\ba\label{SG2fin20}
S^{(2)}_\RM{fin} &=& \RM{tr}_\RM{D}\!\int d^4x\!\int_0^{\infty}ds\,e^{-sm^2}\left[ s^2m^2\SLASH{\omega}(\SLASH{\partial}\SLASH{\omega})\Slash{b}\gamma_5\!+\!2m^2s^3\SLASH{\omega}\SLASH{D}(\partial_\alpha\SLASH{\omega})D^\alpha\Slash{b}\gamma_5\!+\!2m^2s^3\SLASH{\omega}(\partial_\alpha\SLASH{\omega})D^\alpha\SLASH{D}\Slash{b}\gamma_5\right. \nonumber\\
&&\left.-2m^2s^3\SLASH{\omega}\SLASH{\omega}(\partial_\alpha\SLASH{\omega})\omega^\alpha\Slash{b}\gamma_5-2m^2s^3\SLASH{\omega}(\partial_\alpha\SLASH{\omega})\omega^\alpha\SLASH{\omega}\Slash{b}\gamma_5\right]e^{-s\Box}\delta(x-x')\big|_{x'=x},
\ea
where the last two terms do not contribute to the gravitational Chern-Simons term. By performing the trace over matrices and using (\ref{ints}) and (\ref{int}), we obtain
\ba\label{SG2fin2}
S^{(2)}_\RM{CS} &=& \frac i4\int d^4x\int_0^{\infty}ds\,e^{-sm^2} b_\mu\,\omega_{\nu ab}\,\partial_\lambda\omega_{\rho cd} \int\frac{d^4k}{(2\pi)^4}e^{sk^2}s^2m^2\epsilon^{\mu\nu\lambda\rho}(g^{ac}g^{bd}-g^{ad}g^{bc}) \nonumber\\
&=& \frac1{32\pi^2}\int d^4x \epsilon^{\mu\nu\lambda\rho}b_\mu\partial_\nu\omega_{\lambda ab}\,{\omega_\rho}^{ba},
\ea
with $\epsilon^{\mu\nu\lambda\rho}=e\,{e^\mu}_a{e^\nu}_b{e^\lambda}_c{e^\rho}_d\epsilon^{abcd}$. Finally, the relevant $b_\mu\,\omega_{\nu ab}\,{\omega_\lambda}^{bc}\,{\omega_{\rho c}}^a$ finite terms are given by
\ba
S^{(3)}_\RM{fin} &=& i\,\RM{tr}_\RM{D}\! \int d^4x \int_0^{\infty}ds\,e^{-sm^2}\left[\frac{s^2}{2}m^2\SLASH{\omega}\SLASH{\omega}\SLASH{\omega}\Slash{b}\gamma^5+\frac{s^3}{3}m^2\left(\SLASH{\omega}\SLASH{D}\SLASH{\omega}\SLASH{D}\SLASH{\omega}+\SLASH{\omega}\SLASH{D}\SLASH{\omega}\SLASH{\omega}\SLASH{D}+\SLASH{\omega}\SLASH{\omega}\SLASH{D}\SLASH{\omega}\SLASH{D}\right)\Slash{b}\gamma_5\right.\nonumber\\
&&\left.-\frac{s^3}{3}m^2\left(\SLASH{\omega}\SLASH{D}\SLASH{\omega}\SLASH{\omega}+\SLASH{\omega}\SLASH{\omega}\SLASH{D}\SLASH{\omega}+\SLASH{\omega}\SLASH{\omega}\SLASH{\omega}\SLASH{D}\right)b\cdot D\gamma_5-\frac{s^3}{3}m^4\SLASH{\omega}\SLASH{\omega}\SLASH{\omega}\Slash{b}\gamma^5\right]e^{-s\Box}\delta(x-x')\big|_{x'=x}.
\ea
Thus, after calculating the trace the above expression can be written as
\ba
S^{(3)}_\RM{CS} &=& -\frac i{16} \int d^4x\int_0^{\infty}ds\,e^{-sm^2} b_\mu\,\omega_{\nu ab}\,\omega_{\lambda cd}\,\omega_{\rho ef} \\ &&\times\int\frac{d^4k}{(2\pi)^4}e^{sk^2}\left(\frac{s^2}2m^2\epsilon^{\mu\nu\lambda\rho}+\frac{s^3}3m^2\epsilon^{\mu\nu\lambda\rho}k^2+\frac{s^3}3m^2\epsilon^{\alpha\nu\lambda\rho}k_\alpha k^\mu-\frac{s^3}3m^4\epsilon^{\mu\nu\lambda\rho}\right) \nonumber\\
&&\times\left[g^{af}(g^{bc}g^{de}\!-\!g^{bd}g^{ce})\!+\!g^{ae}(g^{bd}g^{cf}\!-\!g^{bc}g^{df})\!+\!g^{ad}(g^{bf}g^{ce}\!-\!g^{be}g^{cf})\!+\!g^{ac}(g^{be}g^{df}\!-\!	g^{bf}g^{de})\right]. \nonumber
\ea
By integrating over the momenta $k$ and the parameter $s$, we have
\be
S^{(3)}_\RM{CS} = -\frac1{48\pi^2}\int d^4x \epsilon^{\mu\nu\lambda\rho} b_\mu\,\omega_{\nu ab}\,{\omega_\lambda}^{bc}\,{\omega_{\rho c}}^a,
\ee
so that combining this expression with (\ref{SG2fin2}) we find the gravitational Chern-Simons term \cite{JacPi} given by  
\be
S_\RM{CS} = \frac{1}{32\pi^2} \int d^4x \epsilon^{\mu\nu\lambda\rho}b_\mu\left(\partial_\nu\omega_{\lambda ab}\,{\omega_\rho}^{ba} - \frac23\omega_{\nu ab}\,{\omega_\lambda}^{bc}\,{\omega_{\rho c}}^a\right).
\ee
This expression can be treated as a four-dimensional analog of the result found in \cite{Oji}. By using the expressions of the vierbein and spin connection in terms of the metric fluctuation $h_{\mu\nu}$, $e_{\mu a} = g_{\mu a} + \frac{1}{2} h_{\mu a}$ and $\omega_{\mu ab} = -\frac{1}{2}\partial_a h_{\mu b} + \frac{1}{2}\partial_b h_{\mu a}$, respectively, we can easily verify that in the weak field approximation this term does not reproduce the value of the numerical coefficient for the linearized gravitational Chern-Simons term obtained in \cite{Mar2}. Comparing with \cite{Mar2}, we note that in the case of the proper-time method, the limit $m^2\to 0$ is not necessary because the divergent contributions vanish.

\section{Summary}

We have applied the proper-time method for the calculation of the Lorentz-breaking Chern-Simons terms in the four-dimensional Yang-Mills and non-linearized gravity theories. These contributions are shown to be finite. For the gravity theory, we did not reproduce the result obtained earlier in \cite{Mar2}, and therefore we can conclude that the gravitational Chern-Simons term has a finite but regularization dependent coefficient, similarly to what happens with the Chern-Simons coefficient in the Lorentz-breaking Yang-Mills theory. It is natural to suggest that, in the case of gravity, the undetermined value of this coefficient is a natural implication of the gravitational triangle anomaly \cite{AlvWit}. The important feature of our result for the gravitational Chern-Simons term is that it is obtained without any restrictions on the field configuration and approximations. At the same time, the result for the Yang-Mills theory is shown to reproduce one of the results obtained in \cite{Gom} in different regularizations scheme. 

{\bf Acknowledgements.} Authors are grateful to Prof.~A.~Das and to Dr.~E.~Passos for useful discussions. A.~Yu.~P. thanks Prof.~R.~Jackiw for some enlightenments. This work was partially supported by Funda\c{c}\~{a}o de Amparo \`{a} Pesquisa do Estado de S\~{a}o Paulo (FAPESP) and Conselho Nacional de Desenvolvimento Cient\'{\i}fico e Tecnol\'{o}gico (CNPq). The work by T.~M. has been supported by FAPESP, project 06/06531-4. The work by A.~Yu.~P. has been supported by CNPq-FAPESQ DCR program, CNPq project No. 350400/2005-9.


\begin{thebibliography}{50}
\bibitem{KosSam} V. A. Kostelecky, S. Samuel, Phys. Rev. D {\bf 39}, 683 (1989).
\bibitem{CarFieJac}S. Carroll, G. Field, R. Jackiw, Phys. Rev. D {\bf 41}, 1231 (1990).
\bibitem{ColKos}D.~Colladay and V.~A.~Kostelecky, Phys.\ Rev.\  D {\bf 55}, 6760 (1997) [hep-ph/9703464]; Phys.\ Rev.\  D {\bf 58}, 116002 (1998) [hep-ph/9809521].
\bibitem{GZK}K.~Greisen, Phys.\ Rev.\ Lett.\  {\bf 16}, 748 (1966); G.~T.~Zatsepin and V.~A.~Kuzmin, JETP Lett.\  {\bf 4}, 78 (1966) [Pisma Zh.\ Eksp.\ Teor.\ Fiz.\  {\bf 4}, 114 (1966)].
\bibitem{Ame} G. Amelino-Camelia, ``The three perspectives on the quantum-gravity problem and their implications for the fate of Lorentz symmetry'' [gr-qc/0309054].
\bibitem{Mag} J. Magueijo, L. Smolin, Class. Quant. Grav. {\bf 21}, 1725 (2004); Phys. Rev. D67, 044017 (2003).
\bibitem{SeiWit}N.~Seiberg and E.~Witten, JHEP {\bf 9909}, 032 (1999) [hep-th/9908142].
\bibitem{CarKos} S. Carroll, V. A. Kostelecky, J. A. Harvey, C. Lane, T. Okamoto, Phys. Rev. Lett. 87, 141601 (2001) [hep-th/0105082].
\bibitem{1}S. Coleman and S. L. Glashow, Phys. Rev. D {\bf59}, 116008 (1999) [hep-ph/9812418].
\bibitem{2}R. Jackiw and V. A. Kosteleck\'y, Phys. Rev. Lett. {\bf82}, 3572 (1999) [hep-ph/9901358].
\bibitem{3}M. P\'erez-Victoria, Phys. Rev. Lett. {\bf83}, 2518 (1999) [hep-th/9905061].
\bibitem{4}J. M. Chung and P. Oh, Phys. Rev. D {\bf60}, 067702 (1999) [hep-th/9812132].
\bibitem{5}J. M. Chung, Phys. Rev. D {\bf60}, 127901 (1999) [hep-th/9904037].
\bibitem{6}W. F. Chen, Phys. Rev. D {\bf60}, 085007 (1999) [hep-th/9903258].
\bibitem{7}J. M. Chung, Phys. Lett. B {\bf461}, 138 (1999) [hep-th/9905095].
\bibitem{8}C.~Adam and F.~R.~Klinkhamer, Phys.\ Lett.\ B {\bf 513}, 245 (2001) [hep-th/0105037].
\bibitem{9}G. Bonneau, Nucl. Phys. B {\bf593}, 398 (2001) [hep-th/0008210].
\bibitem{10}Yu. A. Sitenko, Phys. Lett. B {\bf515}, 414 (2001) [hep-th/0103215].
\bibitem{11}M. Chaichian, W. F. Chen, and R. Gonz\'alez Felipe, Phys. Lett. B {\bf 503}, 215 (2001) [hep-th/0010129].
\bibitem{12}J. M. Chung and B. K. Chung, 105015 (2001) [hep-th/0101097].          
\bibitem{13}A. A. Andrianov, P. Giacconi, and R. Soldati, JHEP {\bf 0202}, 030 (2002) [hep-th/0110279].
\bibitem{14}Yu.~A.~Sitenko and K.~Y.~Rulik, Eur.\ Phys.\ J.\  C {\bf 28}, 405 (2003) [hep-th/0212007].
\bibitem{15}D.~Bazeia, T.~Mariz, J.~R.~Nascimento, E.~Passos and R.~F.~Ribeiro, J.\ Phys.\ A {\bf 36}, 4937 (2003) [hep-th/0303122].
\bibitem{16}Y.~L.~Ma and Y.~L.~Wu, Phys.\ Lett.\  B {\bf647}, 427 (2007) [hep-ph/0611199]. 
\bibitem{Nas}J. R. Nascimento, R. F. Ribeiro and N. F. Svaiter, ``Radiatively induced Lorentz and CPT violation in QED at finite temperature,'' hep-th/0012039.
\bibitem{Cer}L. Cervi, L. Griguolo, and D. Seminara, Phys. Rev. D {\bf64}, 105003 (2001) [hep-th/0104022].
\bibitem{Ebe}D. Ebert, V. Ch. Zhukovsky and A. S. Razumovsky, Phys. Rev. D {\bf 70}, 025003 (2004) [hep-th/0401241].
\bibitem{Mar}T. Mariz, F. A. Brito, J. R. Nascimento, E. Passos and R. F. Ribeiro, JHEP {\bf 0510}, 019 (2005) [hep-th/0509008]. 
\bibitem{Nas2}J.~R.~Nascimento, E.~Passos, A.~Yu.~Petrov and F.~A.~Brito, JHEP {\bf 0706}, 016 (2007), arXiv: 0705.1338 [hep-th].
\bibitem{Gom}M.~Gomes, J.~R.~Nascimento, E.~Passos, A.~Yu.~Petrov and A.~J.~da Silva, Phys.\ Rev.\  D {\bf 76}, 047701 (2007), arXiv: 0704.1104 [hep-th].
\bibitem{Mar2}T.~Mariz, J.~R.~Nascimento, E.~Passos and R.~F.~Ribeiro, Phys.\ Rev.\ D {\bf 70}, 024014 (2004) [hep-th/0403205].
\bibitem{Jac}R.~Jackiw, Int.\ J.\ Mod.\ Phys.\  B {\bf 14}, 2011 (2000) [hep-th/9903044].
\bibitem{Per}M.~Perez-Victoria, JHEP {\bf 0104}, 032 (2001) [hep-th/0102021].
\bibitem{Sch}J.~Schwinger, Phys.\ Rev.\ 82, 664 (1951).
\bibitem{Oji} S. Ojima, Progr. Theor. Phys. 81, 512 (1989).
\bibitem{ColMcD}D.~Colladay and P.~McDonald, Phys.\ Rev.\  D {\bf 75}, 105002 (2007) [hep-ph/0609084].
\bibitem{JacPi} R. Jackiw, S. Y. Pi, Phys. Rev. D {\bf 68}, 104012 (2003) [hep-th/0308071].
\bibitem{AlvWit} L. Alvarez-Gaume, E. Witten, Nucl. Phys. {\bf B234}, 269 (1983).
\bibitem{AlvGin}L. Alvarez-Gaume and P. Ginsparg, Nucl. Phys. {\bf B243}, 449 (1984).
\bibitem{Fuj} K. Fujikawa, Phys. Rev. D {\bf 21}, 2848 (1980), Phys. Rev. Lett. {\bf 42}, 1195 (1979).
\bibitem{FujOji}K. Fujikawa, S. Ojima, S. Yajima, Phys. Rev. D {\bf 34}, 3223 (1986).
\end{thebibliography}
\end{document}